\documentclass[11pt, a4paper]{article} 

\usepackage[top=25mm, bottom=25mm, left=25mm, right=25mm]{geometry}

\usepackage{amsmath}
\usepackage{amsfonts}
\usepackage{amssymb}
\usepackage{bbm,bm}
\usepackage{mathtools}
\usepackage{slashed}
\usepackage{mathalfa}
\usepackage{graphicx,caption,subfigure}
\usepackage{tikz} 
\usepackage{tikz-cd}
\usetikzlibrary{decorations.pathreplacing,calc,tikzmark}
\usepackage{booktabs}
\usepackage{adjustbox}
\usepackage[utf8]{inputenc}
 \usepackage[splitrule]{footmisc}

\usepackage{placeins}
\usepackage{empheq}
\usepackage{paralist}
\usepackage{cite}
\usepackage[normalem]{ulem}
\usepackage{soul}

\usepackage[colorlinks=false,urlbordercolor=red
]{hyperref}
\usepackage{xcolor}
\usepackage{tensor}

\allowdisplaybreaks[2]
\numberwithin{equation}{section}

\DeclareUnicodeCharacter{2212}{-}
\begin{document}

\vspace*{-1.5cm}
\begin{flushright}
  {\small
  LMU-ASC 20/23\\
  MPP-2023-126
  }
\end{flushright}

\vspace{1.5cm}
\begin{center}
  {\Large \bf A note on modular invariant species scale and potentials} 
\vspace{0.35cm}

\end{center}

\vspace{0.35cm}
\begin{center}
{\large 
Niccol\`o Cribiori$^a$ and Dieter L\"ust$^{a,b}$
}
\end{center}

\vspace{0.1cm}
\begin{center} 
\emph{
$^a$Max-Planck-Institut f\"ur Physik (Werner-Heisenberg-Institut), \\[.1cm] 
   F\"ohringer Ring 6,  80805 M\"unchen, Germany, 
   \\[0.1cm] 
 \vspace{0.3cm}
$^b$Arnold-Sommerfeld-Center for Theoretical Physics,\\ Ludwig-Maximilians-Universit\"at, 80333 M\"unchen, Germany, \\[.1cm] 
    } 
\end{center} 

\vspace{0.5cm}

\begin{abstract}

The species scale provides an upper bound for the ultraviolet cutoff of effective theories of gravity coupled to a number of light particle species.
We point out that modular invariant (super-)potentials provide a simple and computable expression of the species scale as a function of the moduli in toroidal orbifold compactifications of type II and heterotic string. Due to modular symmetry, these functions are valid over all moduli space and not only in asymptotic regions. We observe that additive logarithmic corrections to the species scale arise from the requirement that the latter be modular invariant. 
We recast the moduli-dependent expression of the species scale in terms of the gravitino mass or the scalar potential of these models and we connect it to swampland conjectures such as the anti-de Sitter distance conjecture and the gravitino conjecture.

\end{abstract}

\thispagestyle{empty}
\clearpage

\setcounter{tocdepth}{2}

\tableofcontents


\section{Introduction}

Effective theories of gravity typically arise with a number of light scalar fields. In string theory, these can be moduli of the compactification determining various quantities, such as masses, gauge and Yukawa couplings; it is reasonable to expect that they also determine the cutoff scales of the effective description. Indeed, in string theory any parameter is rather a field-dependent function, the primary example being the string coupling, which is determined dynamically as the vacuum expectation value of the dilaton.
In this note, we are interested in the field dependence of the ultraviolet cutoff of effective theories coupled to gravity. 

In the presence of a high number $N_{sp}$ of light fields, or species, it has been proposed that the scale at which gravity becomes strongly coupled is not really the Planck mass, $M_P$, but rather the so-called species scale, $\Lambda_{sp}= M_P /N_{sp}^{\frac{1}{d-2}}$  \cite{Dvali:2007hz,Dvali:2007wp,Dvali:2009ks,Dvali:2010vm,Dvali:2012uq}; see also \cite{Veneziano:2001ah} for earlier work. Here, the adjective light has to be understood with respect to the scale $\Lambda_{sp}$; a high number of light species in the low energy theory affects the ultraviolet cutoff in a direct and non-trivial way. This is a form of UV/IR mixing, which is one of the central themes of the swampland program \cite{Palti:2019pca,Agmon:2022thq}; see \cite{Castellano:2021mmx,Blumenhagen:2022dbo,Cota:2022yjw, Castellano:2022bvr,Cota:2022maf,Castellano:2023qhp,Cribiori:2023ffn, Blumenhagen:2023yws} for recent works in relation to the species scale.

One of the many lessons from string theory is that there are no free parameters. In this respect, it would be hard to embed the number $N_{sp}$ into a consistent effective theory of quantum gravity unless it is described by a field-dependent quantity, $N_{sp} = N_{sp} (\phi)$. In a general setting, it is not clear yet what should govern the functional form of  $N_{sp}(\phi)$. In the specific case of compactifications of the type II string on Calabi-Yau threefolds, an investigation of the moduli dependence of the species scale has been initiated recently in \cite{vandeHeisteeg:2022btw,Cribiori:2022nke,vandeHeisteeg:2023ubh}. These models have an exact moduli space preserving eight supercharges, which allow for computational control over BPS quantities even away from asymptotic regions. Indeed, building on \cite{Long:2021jlv}, in \cite{vandeHeisteeg:2022btw} the function $N_{sp}(\phi)$  has been identified with an index-like quantity, namely the genus-one free energy of the topological string \cite{Bershadsky:1993ta,Bershadsky:1993cx}, receiving contributions from the supersymmetric spectrum of massive excitations. 

In the present note, we look at the field-dependence of the species scale over moduli spaces preserving four supercharges. The associated models typically feature a scalar potential, whose presence can be argued to be fairly general from a swampland perspective \cite{Palti:2020qlc}. As a consequence, scalar fields are typically not exact moduli and regions away from the boundary of field space are harder to reach while maintaining computational control. To overcome this issue, one can follow orbits generated by the action of duality transformations leaving the equations of motion invariant. These orbits extend into the bulk of field space in such a way that the profile of functions like masses or gauge couplings along them can be reliably derived. A prototype example of models in which explicit computations can be performed are six-dimensional toroidal orbifolds of the $E_8 \times E_8$ heterotic string preserving four supercharges, on which we will concentrate.
In this class, duality transformations are in fact modular transformations stemming from the ${\rm PSL}(2,\mathbb{Z})$ duality group of the two-torus. As a consequence, the species scale becomes a modular invariant function of the scalar fields. The logic is nevertheless general and can in principle be applied to more complicated setups. In particular, it can be applied to generic Calabi-Yau manifolds by replacing modularity with mirror symmetry as guiding principle, in case one wants to remain within compactifications of the type II string. The toroidal orbifold case allows us to work with an explicit model at each step of the discussion. 

The study of modular invariant, four-dimensional effective string theory actions with four supercharges dates back to \cite{Ferrara:1989bc,Ferrara:1989qb,Font:1990nt,Dixon:1990pc,Font:1990gx,Cvetic:1991qm,Ferrara:1991uz} and it has recently experienced renewed interest, e.g.~in \cite{Gonzalo:2018guu,Lee:2019tst,Leedom:2022zdm}. We will recall some aspects of these models and give support to the idea that the species scale can be identified with their topological free energy. We will also explain how, when choosing to preserve modularity instead of holomorphy of the species scale, additive logarithmic corrections arise in such a way that $\Lambda_{sp}$ turns out to be slightly larger than the string scale. This is a quantum effect whose mathematical counterpart is the Quillen holomorphic anomaly.

Next, we will connect the species scale to well-known functions of $\mathcal{N}=1$ supergravity in four dimensions, such as the gravitino mass and the scalar potential. Recent works studying the species scales in models with a (positive) potential are \cite{Scalisi:2018eaz,Cribiori:2021gbf,Andriot:2023isc,vandeHeisteeg:2023uxj}. Here, we will not put restrictions on the sign of the potential, which we take to be given by the most general form allowed in $\mathcal{N}=1$ supergravity. Nevertheless, our discussion will be mostly relevant for negative potentials, for which we can make a direct connection to swampland conjectures, such as the anti-de Sitter distance conjecture \cite{Lust:2019zwm} and the gravitino conjecture \cite{Cribiori:2021gbf,Castellano:2021yye}.

Before starting our discussion, two additional comments are in order. First, it has recently been proposed in \cite{Cribiori:2023ffn} that $N_{sp}$ should really be understood as an entropy of species, $\mathcal{S}_{sp}$. We will take this proposal seriously in the present work and thus all the results presented below should be understood from this point of view as results on $\mathcal{S}_{sp}$.
Second, we will not keep track of many order one factors in our analysis. The coefficients we display explicitly are those needed for self-consistency among the various formulae we will use. 

This work is organized as follows. In section \ref{sec:freeenergy}, we review and motivate further the relation between the species scale and the topological free energy, from both a black hole perspective and a direct computation in the specific toroidal orbifold of interest. 
In section \ref{sec:modinvlogcorr}, we study the role of modular invariance and explain how this is related to the presence of additive logarithmic corrections to the species scale. 
In section \ref{sec:potential}, we connect our discussion to (modular invariant) superpotential and scalar potentials of $\mathcal{N}=1$ supergravity in four dimensions; we comment on how the anti-de Sitter distance conjecture \cite{Lust:2019zwm} and the gravitino conjecture are automatically realized in our setups \cite{Cribiori:2021gbf,Castellano:2021yye}. In section \ref{sec:discussion}, we draw our conclusions and point out possible future directions of research.

\section{Species scale and free energy}
\label{sec:freeenergy}

The species scale has been proposed as a well-defined ultraviolet cutoff for the gravitational sector of a given effective theory \cite{Dvali:2007hz,Dvali:2007wp,Dvali:2009ks,Dvali:2010vm,Dvali:2012uq}, see also \cite{Veneziano:2001ah} for earlier work. For a $d$-dimensional theory of gravity coupled to a number $N_{sp}$ of light species, it reads
\begin{equation}
\label{Lambdasp}
\Lambda_{sp} = \frac{M_P}{N_{sp}^{\frac{1}{d-2}}}\, .
\end{equation}
As such, when $N_{sp}$ is high $\Lambda_{sp}$ deviates sensibly from the Planck mass $M_P$, namely the naive estimate for the quantum gravity cutoff.

The proposal can be motivated from both perturbative and non-perturbative considerations. 
In the first case, one studies corrections to the graviton propagator due to loops of $N_{sp}$ particles. In the worst scenario, the perturbative expansion breaks down at the scale in which the first order correction becomes comparable to the tree level term. This scale is indeed \eqref{Lambdasp}.\footnote{For massless particles running in the loop, one actually finds $N_{sp}\log N_{sp}$ instead of just $N_{sp}$. We will not discuss this effect in the present work.} 
In the second case, $\Lambda_{sp}$ provides a lower bound for the entropy (and thus the size) of the minimal black hole that can be reliably described within the effective theory.  For a black hole of radius $R_{BH}$, the semiclassical entropy scales as $\mathcal{S}\sim (R_{BH}M_p)^{d-2}$. Naively, one would expect the minimal entropy to be of order one, corresponding to a black hole of planckian size, $R_{BH}\sim 1/M_P$. However, in a theory with $N_{sp}$ species this leads to a contradiction, for in principle the black hole should be able to emit or absorb all particle species. Thus, the minimal entropy has to scale as $\mathcal{S}_{min}\sim N_{sp}$ and one finds instead $R_{BH} \sim 1/\Lambda_{sp}$ as the radius of the smallest possible black hole. 

This second argument connects the number of species to the black hole entropy. A more radical point of view, which we adopt in the present of work, is to think of $N_{sp}$ as an entropy of species, $\mathcal{S}_{sp}$, with an associated thermodynamics, as recently proposed in \cite{Cribiori:2023ffn}. In what follows, we identify systematically $\mathcal{S}_{sp} \equiv N_{sp}$ and our results have to be interpreted along with this logic.

Given that the value of the ultraviolet cutoff typically depends on the spectrum of the given effective theory, the relation \eqref{Lambdasp} is in some sense non-linear. This makes computing $\Lambda_{sp}$ non-trivial in general. Recently, a moduli-dependent expression for the species scale in string compactifications has been proposed in \cite{vandeHeisteeg:2022btw} and subsequently motivated in \cite{Cribiori:2022nke} from the entropy of small extremal black holes. In the present section, we review these works together with further developments in connection to the distance conjecture \cite{vandeHeisteeg:2023ubh}. 
In addition, we elaborate on the relation between the proposed moduli-dependent $\Lambda_{sp}$ and the topological free energy as originally defined in \cite{Ooguri:1991fp,Ferrara:1991uz}, which in turn is closely related to the moduli-dependence of threshold corrections derived in the seminal work \cite{Dixon:1990pc}.

\subsection{Moduli-dependent species scale in type II compactifications}

In string theory, one expects $\Lambda_{sp}$ to be a function of the moduli of the compactification, as it is the case for other scales and couplings. In \cite{vandeHeisteeg:2022btw}, it has been proposed that the dependence of the species scale on the vector multiplet moduli space of type II compactifications on a Calabi-Yau threefold is captured by 
\begin{equation}
\label{Nsp=F1}
N_{sp} \simeq F_1,
\end{equation}
where $F_1$ is the genus-one free energy of the topological string with target space the same threefold \cite{Bershadsky:1993ta,Bershadsky:1993cx}. One can write it in a closed form as \cite{Bershadsky:1993ta}
\begin{equation}
\label{F1topstring}
F_1 = \log Q(t,\bar t) f(t) \bar f(\bar t),
\end{equation}
where $f(t)$ is an holomorphic function of the moduli, to be fixed by boundary conditions on the moduli space, while $Q(t,\bar t)$ is a non-holomorphic piece which can be understood as a Quillen anomaly \cite{Bershadsky:1993cx}. For the topological A-model, related to the K\"ahler moduli space of type IIA compactifications, one has
\begin{equation}
Q(t, \bar t) = \exp\left(\frac12\left( h^{1,1}+3 +\frac{\chi}{12}\right)K\right)\left({\det g_{i\bar \jmath}}\right)^{-\frac12},
\end{equation}
where $\chi=2(h^{2,1}-h^{1,1})$ is the Euler characteristic of the threefold, while $K$ and $g_{i\bar\jmath}$ the K\"ahler potential and metric respectively. The expression for the topological B-model, related to the complex structure moduli space of type IIB compactifications, can be found by replacing $h^{1,1}\to h^{2,1}$ and hence $\chi \to -\chi$.

An argument in favour of \eqref{Nsp=F1} and based on the entropy of the smallest possible black hole in the effective theory has been provided in \cite{Cribiori:2022nke}. The technique employed there to study limits of small (or large) entropy engineered from limits on the moduli space has been developed in \cite{Bonnefoy:2019nzv,Cribiori:2022cho,Delgado:2022dkz} and it can be applied in fairly general configurations. Instead of giving an overall presentation, we prefer to proceed by examples, from which the broader picture can be deduced. 

Consider an extremal black hole obtained from $q$ $D0$-branes and $p^i$ $D4$-branes wrapped on holomorphic four-cycles of a Calabi-Yau threefold on which type IIA string theory is compactified. Equivalently, consider M-theory on the same Calabi-Yau, with $p^i$ $M5$-branes wrapped on holomorphic four-cycles, times a circle with $q$ units of momentum. The microscopic entropy of the setup has been originally computed in the seminal work \cite{Maldacena:1997de} and successfully matched with the macroscopic supergravity quantity in \cite{LopesCardoso:1998tkj}.
Before recalling the derivation of \eqref{Nsp=F1} given in \cite{Cribiori:2022nke}, let us check that the strategy leads to a sensible answer in a simpler example.

In the large volume limit, the dynamics on the K\"ahler moduli space of type IIA compactifications is governed by the classical prepotential
\begin{equation}
F_0(X) = -\frac16 C_{ijk}\frac{X^i X^j X^k}{X^0},
\end{equation}
where the symplectic sections $X^\Lambda=X^\Lambda(z)$, with $\Lambda=\{0,i\}$, depend on the $i=1,\dots,h^{1,1}$ moduli $z^i = X^i/X^0$. The coefficients $C_{ijk}$ are the intersection numbers of the Calabi-Yau. The masses of the BPS states are proportional to the covariantly holomorphic sections $e^{\frac K2} X^\Lambda$ and $e^{\frac K2} \partial_\Lambda F_0$, where $K$ is again the K\"ahler potential. The symplectic invariant linear combination of BPS masses is the central charge of the supersymmetry algebra,
\begin{equation}
Z = e^\frac{K}{2}\left(q_\Lambda X^\Lambda - p^\Lambda \partial_\Lambda F_0\right) .
\end{equation}
To match with the microscopic model of \cite{Maldacena:1997de}, we choose non-vanishing charges $-q_0=q>0$ and $p^i>0$,
supporting a BPS black hole solution. 
The attractor equations \cite{Ferrara:1996dd}
\begin{equation}
p^\Lambda = i \left(X^\Lambda - \bar X^\Lambda\right), \qquad q_\Lambda = i \left(\partial_\Lambda F_0 - \partial_\Lambda \bar F_0\right)
\end{equation}
fix the moduli at the black hole horizon in terms of the charges and allow to compute the entropy as
\begin{equation}
\mathcal{S} = \pi Z \bar Z = 2 \pi \sqrt{\frac q6 C_{ijk}p^ip^jp^k}.
\end{equation}
At the same time, the Calabi-Yau volume modulus $\mathcal{V}_6 = \frac16 C_{ijk}t^i t^j t^k$, with $t^i ={\rm Im}z^i$, is given at the horizon by
\begin{equation}
\mathcal{V}_6 = \sqrt{\frac{q^3}{\frac16 C_{ijk}p^ip^jp^k}}.
\end{equation}
Since charges are quantized, the minimal non-vanishing entropy in this setup is obtained for a charge configuration such that $\frac16 C_{ijk}p^ip^jp^k=1$, leading thus to a lower bound 
\begin{equation}
\label{SVol13}
\mathcal{S}>\mathcal{S}_{min} = 2 \pi q^{\frac12} = 2 \pi \mathcal{V}_6^\frac13 \simeq N_{sp} \equiv \mathcal{S}_{sp}.
\end{equation}
We find that the entropy of the smallest black hole which can be reliably described by the effective theory is given by the volume of a two-cycle, which indeed scales as $\mathcal{V}_6^\frac13$, if $\mathcal{V}_6$ is the volume of the whole threefold. We interpret this result as the fact that the minimal entropy detects the smallest topologically non-trivial cycle in the manifold. For simply-connected Calabi-Yau threefolds, it cannot be a one-cycle. Turning the logic around, one could have learned that the underlying geometry is simply-connected from the requirement that the minimal entropy gives the volume of the minimal cycle. The appearance of the power $\mathcal{V}_6^\frac13 $ can be understood more concretely. Indeed, from direct implementation of the attractor mechanism one finds
\begin{equation}
\label{SVolX0X0b}
\mathcal{S}=\pi Z \bar Z=\pi e^{-K} = 8\pi\, \mathcal{V}_6 X^0 \bar X^0, 
\end{equation}
where $X^0$ is the additional scalar of the homogeneous coordinates. Since at the horizon $X^0 = -\frac12 \sqrt{\frac{\frac 16 C_{ijk}p^ip^jp^k}{q}}$, for $\frac16 C_{ijk}p^ip^jp^k=1$ we have $X^0 \simeq \mathcal{V}_6^{-\frac13}$ and thus \eqref{SVolX0X0b} reduces to \eqref{SVol13}.
This result is consistent with the expression of the species scale $\Lambda_{sp} \simeq (N_{sp})^\frac{1}{k} M_{KK}$, valid when the species are Kaluza-Klein states associated to $k$-compact dimensions. Indeed, from the black hole argument we learn that the species arise from $k=2$ compact dimensions. Since for a two-dimensional compactification one has $M_{KK} \simeq M_s/\mathcal{V}_6^\frac16 \simeq M_P /\mathcal{V}_6^{\frac13}$, the relation $\Lambda_{sp} \simeq (N_{sp})^\frac{1}{k} M_{KK} \simeq M_P/\sqrt{N_{sp}}$ is solved precisely by \eqref{SVol13} (we are neglecting the dependence on the string coupling, which instead governs string states).

Now, we would like to recover \eqref{Nsp=F1} as in \cite{Cribiori:2022nke}. To this purpose, we need to supplement the supergravity effective action with higher derivative corrections. Indeed, it is known from \cite{Antoniadis:1993ze} that the topological string computes specific higher derivative corrections to four-dimensional $\mathcal{N}=2$ supergravity.\footnote{There is a caveat to this statement, see \cite{Cardoso:2008fr,Cardoso:2010gc,Cardoso:2014kwa}, to which we will come back in the conclusions.} 
In particular, the genus-one free energy couples to the Gauss-Bonnet term in a way that resembles a gravitational theta-term,
\begin{equation}
\label{RwsR}
\mathcal{L}_{corr} \sim F_1 R \wedge * R.
\end{equation}
In order not to break supersymmetry at the Lagrangian level, one has to interpret $F_1$ as a correction to the classical prepotential $F_0$, leading to a generalized prepotential
\begin{equation}
\label{F0F1A}
F(X,A) = F_0(X) + F_1(X) A.
\end{equation}
On the other hand, $R \wedge * R$ has to be embedded into a chiral multiplet $A$, usually called graviphoton background multiplet, whose lowest component is the square of the auxiliary $T$-field of the Weyl multiplet, while the highest component is indeed $R \wedge * R$. In the large volume regime in which the supergravity approximation is reliable, one has
\begin{equation}
\label{F1bh}
F_1 = d_i \frac{X^i}{X^0}, \qquad d_i = - \frac{1}{24}\frac{1}{64} c_{2i}, \qquad c_{2i} = \int c_2 \wedge \omega_i,
\end{equation}
where $c_2$ is the second Chern class of the tangent bundle of Calabi-Yau and $\omega_i$ a basis of its two-form cohomology. 
One can then proceed and study black hole solutions in a supergravity theory with prepotential \eqref{F0F1A}.
In practice, one can use the same rules from special geometry and also the same attractor equations, upon replacing $F_0(X) \to F(X,A)$ \cite{Behrndt:1996jn}. In appropriate conventions, the value of the background $A$ at the horizon is fixed to be $A=-64$ \cite{LopesCardoso:1998tkj}.

As is well-known, in the presence of higher derivative corrections the Bekenstein-Hawking formula does not capture the full contribution to the entropy. Using instead the Wald formula \cite{Wald:1993nt}, in \cite{LopesCardoso:1998tkj} the entropy of BPS black holes has been calculated in supergravity for a generic $F(X,A)$, resulting in the simple expression
\begin{equation}
\mathcal{S} = \pi \left[Z \bar Z -256 \, {\rm Im} \,\partial_A F(X,A) \right].
\end{equation}
Notice that also the central charge is modified such that $Z=Z(X,A)$.
For a black hole with only $q$ and $p^i$ non-vanishing charges, this formula successfully reproduces the microscopic result of \cite{Maldacena:1997de}, namely
\begin{equation}
\mathcal{S} = 2 \pi \sqrt{\frac q6 \left(C_{ijk}p^ip^jp^k + c_{2i}p^i\right)} .
\end{equation}
The minimal entropy can now be engineered for a charge configuration which is such that $\frac16 C_{ijk}p^ip^jp^k=0$ while $c_{2i}p^i \neq 0$, giving a lower bound
\begin{equation}
\mathcal{S} >\mathcal{S}_{min} = 2 \pi \sqrt{\frac q6c_{2i }p^i} \simeq F_1 \simeq N_{sp}\equiv \mathcal{S}_{sp}.
\end{equation}
Here, we used that at the horizon the value of $F_1$ in \eqref{F1bh} and with the aforementioned charge configuration is $F_1 \simeq c_{2i}t^i\simeq \sqrt{q c_{2i}p^i}$.\footnote{We are neglecting a factor $i$ which appears when comparing $F_1$ between supergravity and topological string conventions.} 
Thus, the lower bound on the entropy reproduces \eqref{Nsp=F1}, in the appropriate regime of validity of the supergravity approximation.

The relation between species scale and higher derivative corrections to the effective action holds more in general than the specific example considered above, see e.g.~\cite{Arfaei:2023hml} for recent work. This is because the species scale gives the ultraviolet cutoff of the gravity sector of generic effective theories in $d$ dimensions. 
Indeed, consider an higher derivative expansion of the schematic form
\begin{equation}
\mathcal{L} \sim M_p^{d-2}\left(R + \frac{N}{M_p^{d-2}} R^2 +\dots\right),
\end{equation}
where $N$ is a coefficient which can depend on the fields in the theory. Here, we denoted generically with $R^2$ a correction involving two Riemann tensors contracted or traced in any possible way, or a general linear combination thereof. The correction \eqref{RwsR} is a particular sub-case of this, with $N=F_1$. The perturbative expansion definitely breaks down when the first order term, $R^2$, is comparable to the tree level, $R$. Since $R$ has mass dimension two, $R\sim \Lambda^2$, while $N$ has mass dimension $d-4$, $N \sim \Lambda^{d-4}$, it is convenient to introduce the dimensionless quantities $R_{(0)} \sim R/\Lambda^2$, $N_{(0)} \sim N /\Lambda^{d-4}$ and rewrite
\begin{equation}
\mathcal{L} \sim M_p^{d-2} \Lambda^2 \left(R_{(0)}+\frac{N_{(0)}}{M_p^{d-2}}\Lambda^{d-2} (R_{(0)})^2 +\dots\right).
\end{equation}
Then, the perturbative expansion breaks down when the coefficient of the second term is of order one, which happens at a scale
\begin{equation}
\Lambda = \frac{M_p}{N_{(0)}^{\frac{1}{d-2}}} \equiv \Lambda_{sp}.
\end{equation}
We thus recover \eqref{Nsp=F1} as a sub-case associated to specific higher derivative corrections.
This simple argument exemplifies the close relationship between the species scale and gravitational threshold corrections. In turn, in various explicit examples \cite{Antoniadis:1991fh,Antoniadis:1992rq,Antoniadis:1992sa} the latter have a field dependence similar to gauge threshold corrections \cite{Dixon:1990pc}.

\subsection{Relation to swampland conjectures}
\label{sec:dc}

Within the swampland program, a generic expectation is that an infinite tower of states becomes exponentially light when approaching the boundary of moduli space, in accordance with the distance conjecture. For a canonically normalized modulus $\phi$, say with moduli space $\mathbb{R}$, in the asymptotic limit $\phi \to \infty$ the mass of the predicted tower is conjectured to behave as
\begin{equation}
m\sim M_P e^{-\alpha \phi\, M_P^{(2-d)/2}},
\end{equation}
where $\alpha$ is some parameter of order one. The species scale is influenced by the number of light species in the theory.
As a consequence, it is natural to expect a relation such as 
\begin{equation}
\label{LambdaspDC}
\Lambda_{sp} \sim M_P \left(\frac{m}{M_P}\right)^{n},
\end{equation}
as we approach the boundary of moduli space. Here, $n$ is another order one constant parameter which can be model-dependent.

The emergent string conjecture \cite{Lee:2019wij} proposes that there are in fact only two possible towers in quantum gravity, namely either towers of string or of Kaluza-Klein states. To the former, one can associate $\alpha_{string} = \frac{1}{\sqrt{d-2}}$, while to the latter $\alpha_{KK} =\sqrt{\frac{(d+k-2)}{k(d-2)}}$, in the case of a decompactification from $d$ to $d+k$ dimensions; see e.g.~\cite{Agmon:2022thq} for a derivation. Since $\alpha_{KK}>\alpha_{string}$, assuming the emergent string conjecture one can argue that for $\phi\to\infty$ the mass of a generic tower should behave as \cite{Etheredge:2022opl}
\begin{equation}
m \leq M_P e^{-\frac{\phi}{\sqrt{d-2}}M_P^{(2-d)/2}}.
\end{equation}
If the fastest possible decay is exponential, from the emergence string conjecture one can argue for $\Lambda_{sp} \geq M_P e^{-\frac{\phi}{\sqrt{d-2}}M_P^{(2-d)/2}}$. 
Then,  at the boundary of the moduli space one expects the upper bound \cite{vandeHeisteeg:2023ubh}
\begin{equation}
\label{dLambdaspbound}
\left|\frac{\Lambda_{sp}'(\phi)}{\Lambda_{sp}(\phi)} \right|^2 \leq \frac{c_{sp}}{M_P^{d-2}},
\end{equation}
with $c_{sp}$ an order one parameter possibly depending on the number of spacetime dimensions. Here and in the following, derivatives with respect to canonically normalized scalars are denoted with primes. 

Indeed, in \cite{vandeHeisteeg:2023ubh} such a bound is derived from more general considerations and proposed to be valid over all moduli space, not only in the asymptotic regions. The argument of \cite{vandeHeisteeg:2023ubh} exploits the fact that integrating out one (or few) massive mode(s) up to the scale $\Lambda_{sp}$ should not modify drastically the form of the effective action obtained from integrating out all (infinite) modes with mass above $\Lambda_{sp}$.
This reasoning does not really fix $c_{sp}$ to $\frac{1}{d-2}$ and indeed examples discussed in \cite{vandeHeisteeg:2023ubh} approach this constant from above.

\subsection{Topological free energy and toroidal orbifolds}
\label{sec:Ftororb}

Historically, the one-loop topological free energy has been introduced in relation to string theory first in \cite{Ooguri:1991fp,Ferrara:1991uz}, which we follow in the present section. However, a similar function of the moduli, though with a different prefactor, appeared already in \cite{Dixon:1990pc}, where it was understood that it captured the moduli-dependence of threshold corrections to gauge and gravitational couplings. See \cite{Antoniadis:1991fh,Antoniadis:1992rq,Antoniadis:1992sa,LopesCardoso:1994ik,LopesCardoso:1995qa,deWit:1996wq} for related works and developments on orbifold compactifications. 
From a string theory perspective, the topological free energy is given in terms of the Ray-Singer torsion, which is indeed a topological invariant.

From a field theory perspective, the free energy $F$ is defined as the functional integral over the complete mass spectrum. The topological free energy is then the sub-case in which the functional integration is extended only over massive states associated to the topology of the compactification, while omitting the modes present also in the decompactified theory, such as massive oscillator modes. Concretely, one has
 \begin{equation}
e^{F} \simeq \int [d\phi] \,e^{-\frac12 \phi^T \mathcal{M}^2 \phi} \simeq (\det \mathcal{M}^2)^{-\frac12},
\end{equation}
where we are omitting constant factors in the integration and we are neglecting derivative terms in general.\footnote{The kinetic term of the moduli $\phi$ would contribute as constant factors (powers of $2\pi$), while higher derivatives are suppressed by a mass scale assumed to be large.} Depending on which states are included in the mass formula $\mathcal{M}^2$, one gets different definitions of free energy. 
For example, when performing a dimensional reduction of type II or heterotic string on a Calabi-Yau manifold, one expands all fields in harmonic forms of the compact space. These forms are counted by Betti numbers, which are topological invariants. Each of them is associated with a tower of massive states which contributes to the topological free energy.

With the help of supersymmetry, one can work with explicit formulae. In a supersymmetric spectrum, the bosonic and fermionic free energy differ by a sign. 
We can write the fermionic free energy as
\begin{equation}
F \simeq  \log \det M^\dagger M,
\end{equation}
where $M_{\alpha\beta}$ is the fermionic mass matrix.  
In theories with four supercharges, assuming that contributions from D-terms are vanishing, the fermionic mass matrix is given by \cite{Wess:1992cp}
\begin{equation}
\label{Malphabeta}
M_{\alpha\beta} = e^{\frac G2} \left(G_{\alpha \beta} + G_\alpha G_\beta-\Gamma^{\gamma}_{\alpha\beta}G_\gamma\right), 
\end{equation}
where we introduced the K\"ahler invariant combination $G$ of the K\"ahler potential $K$ and superpotential $W$, namely
\begin{equation}
G = K + \log W \bar W,
\end{equation}
and indices attached to it denote flat derivatives. Notice that we are integrating out infinite towers of states arising from the compactification. Hence, we are formally dealing with an infinite number of fields, labelled by $\alpha=1,\dots,\infty$, and the mass matrix \eqref{Malphabeta} is infinite dimensional. Next, in order to take into account the fact that massive fermions have in general a non-canonical kinetic term, we have to multiply by appropriate powers of the K\"ahler metric and thus we rescale $M_{\alpha\beta}\to \tilde M_{\alpha\beta} =\sqrt{G_{\alpha\bar \alpha}} \bar M^{\bar \alpha \bar \beta}\sqrt{G_{\bar \beta \beta}}$. We perform this step right away and omit the tilde in what follows. 
In addition, on a supersymmetric vacuum, $G_\alpha=0$. Taking all of this into account, we arrive at
\begin{equation}
F\simeq \log \det M^\dagger M =\log\left[ \det\left(e^K  (G_{\alpha\bar\beta})^{-2} \right) | \det W_{\alpha\beta}|^2 \right].
\end{equation}
Notice that this expression has the same structure as \eqref{F1topstring}, namely there is an holomorphic contribution, $\det W_{\alpha\beta}$, but also non-holomorphic one, $\det\left(e^K  (G_{\alpha\bar\beta})^{-2}\right) $, arising from the factor $e^\frac{G}{2}$  in \eqref{Malphabeta} and from the rescaling to get canonically normalized kinetic terms.

We focus now on toroidal orbifolds  of the $E_8\times E_8$ heterotic string leading to four-dimensional $\mathcal{N}=1$ theories. A summary of the web of string dualities relating them to other models can be found in \cite{Gonzalo:2018guu}, with a particular eye towards threshold corrections. 
As computed explicitly in \cite{Ferrara:1991uz}, for this class of models one can effectively replace $\det\left(e^K\left(G_{\alpha\bar\beta}\right)^{-2}\right)$ with $\det\left(e^K \delta_{\alpha\bar\beta}\right)$, with $\delta_{\alpha\bar\beta}$ the infinite dimensional diagonal matrix, and thus we find\footnote{More precisely, this is allowed as long as one looks at the untwisted sector of symmetric orbifold models, as we are doing in the present work. In this case, the metric $G_{\alpha\bar\beta}$ is block-diagonal with three infinite dimensional blocks associated to the three two-tori inside the six torus. Each block is in turn diagonal and differs from the identity by a term proportional to the volume of the associated two-torus. This property leads to the simplification mentioned above.}
\begin{equation}
\label{F1Torbifolds}
F\simeq \log\det \left(e^K|W_{\alpha\beta}|^2\right) =  \,{\rm Tr}\log  \left(e^K|W_{\alpha\beta}|^2\right).
\end{equation}
This is the topological free energy of the class of models we are interested in. Its modular invariance makes it an exact result valid over all moduli space.

Let us present the explicit expression for  \eqref{F1Torbifolds} in the case of the $E_8\times E_8$ heterotic string on the orbifold $T^6/\left(\mathbb{Z}_3\times \mathbb{Z}_3\right)$, which will be our main example. Other orbifolds can be considered as well, see e.g.~\cite{Ferrara:1991uz}. This $\mathcal{N}=1$ toroidal orbifold has $h^{11}=3$ and $h^{21}=0$, with moduli space $\left({\rm SU}(1,1)/{\rm U}(1)\right)^{1+h^{11}}$, where the additional factor corresponds to the axio-dilaton. As such, this moduli space captures the dependence on the diagonal moduli of the six-torus and thus one can just perform the computation for three copies of a two-torus.

Following \cite{LopesCardoso:1995qa}, we look at BPS states whose masses can be computed explicitly in terms of integrals over supersymmetric cycles.
For the two-torus, these are the periods $(1,T)$. The most generic BPS mass is thus given by the linear combination $(m+nT)$ with arbitrary integers $m,n\in\mathbb{Z}$. From \cite{LopesCardoso:1995qa}, we are instructed to calculate the free energy by summing over all possible BPS masses. Recalling that the K\"ahler potential for a single two-torus modulus is $K=-\log(-i(T-\bar T))$, the trace over the logarithm of the fermionic mass matrix in \eqref{F1Torbifolds} can be written explicitly as \cite{Ferrara:1991uz}
\begin{equation}
\label{FT6gen}
F \simeq  \sum_{i=1}^{h^{11}}\sum_{(m,n)\neq(0,0)}\log  \frac{|m_i+n_iT_i|^2}{-i(T_i-\bar T_i)} = \sum_{i=1,2,3} \log\left[-i(T_i-\bar T_i)|\eta(T_i)|^4\right],
\end{equation}
where we introduced an index $i=1,2,3\equiv h^{11}$ to distinguish the three two-tori contributions. Notice that from the trace over the BPS masses we omitted the contribution with $m=n=0$, corresponding to a strictly massless state. This would give an infrared divergence, which has thus been already regularized. The infinite sum in \eqref{F1Torbifolds} is also divergent in the ultraviolet, but this can be taken care of with Riemann zeta function regularization. The essential tools are summarized in appendix \ref{app:conv}. For the case of an isotropic torus, $T_1=T_2=T_3\equiv T$, the free energy reduces to the simple expression
\begin{equation}
\label{FT6iso}
F \simeq 3 \log\left[-i(T-\bar T)|\eta(T)|^4\right].
\end{equation}
One can check that, up to an overall factor of $2$, this matches with \eqref{F1topstring} for $K=-3\log (-i(T-\bar T))$, $h^{11}=3$ and $\chi=0$; moreover by direct comparison one finds $f(T) =\eta(T)^{12}$.

\section{Modular-invariant species scale and log-corrections}
\label{sec:modinvlogcorr}

 In this section, we establish the connection between the topological free energy  \eqref{F1Torbifolds}  and the species scale of heterotic toroidal orbifolds. We do so empirically, namely we reconstruct the functions \eqref{FT6gen}, \eqref{FT6iso} by starting from an expression of $\mathcal{S}_{sp}$ valid in the asymptotic regions of moduli space and turning it into a modular invariant quantity. The resulting function will capture the field dependence of the species scale everywhere on moduli space. Concretely, we will be looking at two kinds of moduli, namely the (isotropic) volume modulus and the axio-dilaton. They correspond to the two only possible towers of states according to the Emergent String Conjecture \cite{Lee:2019wij}, respectively Kaluza-Klein and string states. One can generalize the picture along any preferred direction. The results of this section have some overlap with the recent work \cite{vandeHeisteeg:2023ubh}, but their derivation is complementary.

The starting point is that, in first approximation, the species scale is the string scale. We will then find a correction which will allow us to give a physical meaning to the Quillen anomaly in the context of the species scale. 
Imposing that $\Lambda_{sp} \simeq M_s$ and recalling that $M_s \simeq M_P\mathcal{V}_k^{-\frac{1}{d-2}} g_s^{\frac{2}{d-2}}$, where $\mathcal{V}_k$ is the dimensionless volume modulus of the compact $k$-dimensional manifold, $d$ is the number of non-compact dimensions and $g_s$ the string coupling, we find the moduli-dependent expression for the entropy of species
\begin{equation}
\label{Nsp=volgs}
\mathcal{S}_{sp}\simeq \mathcal{V}_k\,g_s^{-2}.
\end{equation}
Since we implicitly used a spacetime effective action to derive it, the above expression is valid at the asymptotic boundary of moduli space, namely at large volume and weak string coupling, where indeed $\mathcal{S}_{sp}$ is large. The function \eqref{Nsp=volgs} depends explicitly on the moduli governing the two towers predicted by the Emergent String Conjecture \cite{Lee:2019wij}. We discuss them separately in the following two sections.

\subsection{Kaluza-Klein states}

Towers of Kaluza-Klein states are governed by the volume of the compact manifold. 
In the asymptotic regions of its moduli space, the volume of a two-torus is given by $\mathcal{V}_2 \simeq -i(T-\bar T) \simeq  {\rm Im}T$. Therefore, for the six-torus we have  
\begin{equation}
\label{NspT6bdr}
\mathcal{S}_{sp}\simeq \mathcal{V}_6^\frac13 \simeq \left( \prod_{i=1}^3 (-i(T_i-\bar T_i))\right)^{\frac13},
\end{equation} 
where we used the result \eqref{SVol13}, namely $\mathcal{V}_6^\frac13 \simeq \mathcal{V}_2$.  
This function captures the moduli dependence of $\mathcal{S}_{sp}$ at the boundary of moduli space. To go beyond this approximation, one has to think of \eqref{NspT6bdr} as the limit of some yet unknown function defined over all moduli space and invariant under the duality group, which is constituted by three copies of ${\rm PSL}(2,\mathbb{Z})$ in this simple example. For this to be realized, there should exist a function of ${\rm Im} \,T$ with the same (divergent) behaviour both at large and small values of its argument. As recalled in appendix \ref{app:conv}, the Dedekind eta function has the property that
\begin{equation}
\eta(T) \sim e^{-\frac{\pi}{12}{\rm Im}T},
\end{equation}
for ${\rm Im}T\to 0$ as well as ${\rm Im}T \to \infty$. Therefore, the educated guess is that 
\begin{equation}
\label{NspT2bu}
\mathcal{S}_{sp} \simeq \log |\eta(T)|^a,
\end{equation}
for the two-torus and 
\begin{equation}
\label{NspT6bu}
\mathcal{S}_{sp} \simeq \sum_{i=1}^3\log |\eta(T_i)|^{a_i},
\end{equation}
for the six-torus. The parameters $a_i$ are fixed to $-4/\pi$ by requiring that one recovers \eqref{NspT6bdr} at the (isotropic) boundary. This is analogous to how the function $f(t)$ in \eqref{F1topstring} is fixed by boundary conditions. As in the previous sections, we will be cavalier on factors of $\pi$ and drop them in what follows. The minus sign instead has a physical meaning and it indicates that the expression above will eventually represent the bosonic free energy, differing by an overall sign from the fermionic one calculated in section \ref{sec:Ftororb}.

We are thus on the right track towards \eqref{FT6gen}, but we still miss the non-holomorphic contribution. Indeed, the expression of $\mathcal{S}_{sp}$ above is not yet modular invariant: \eqref{NspT2bu} is invariant under $T\to T+1$, but it transforms under $T\to -1/T$ as 
\begin{equation}
|\eta(T)|^2 \to \sqrt{T\bar T} \,|\eta(T)|^2.
\end{equation}
As is well-known, one can easily fix this lack of invariance by replacing the quantity $|\eta(T)|^2 $ with $\left(-i(T-\bar T)\right)^\frac12 |\eta(T)|^2$. We thus arrive at the modular invariant expression
\begin{equation}
\label{NspT22bu}
\mathcal{S}_{sp} \simeq -\log \left[-i(T-\bar T)|\eta(T)|^4\right],
\end{equation}
for the two-torus and
\begin{equation}
\label{NspT62bu}
\mathcal{S}_{sp} \simeq -\sum_{i=1}^3 \log \left[-i(T_i-\bar T_i)|\eta(T_i)|^4\right],
\end{equation}
for the six torus. Up to an overall minus sign whose physical meaning has been explained above, we recovered \eqref{FT6gen}.  
Besides, we learned that the Quillen anomaly in \eqref{F1Torbifolds} has the meaning of restoring modular invariance of the species scale. Due to its invariance under modular transformations, the expression \eqref{NspT62bu} is valid over all moduli space, even in regions in which the number of species is of order one. Notice that, as long as modularity is the guiding principle, one can include in \eqref{NspT2bu} and \eqref{NspT6bu} any additional modular invariant function, such as the $j$-invariant $j(T)$ defined in \eqref{jinv}. We will come back to this in section \ref{sec:potential}.

We can now see that the non-holomorphic contribution gives a correction to the entropy (or number) of species such that the species scale at the boundary of the moduli space is slightly bigger than the string scale, which was our approximated starting point. Indeed, considering an isotropic six-torus for simplicity, we have 
\begin{equation}
\mathcal{S}_{sp} \simeq -\log \left[(-i(T-\bar T))^3|\eta(T)|^{12}\right]
\end{equation}
and, for ${\rm Im}T\to \infty$, this behaves as
\begin{equation}
\begin{aligned}
\mathcal{S}_{sp} \simeq \mathcal{V}_6^\frac13 - 3 \log \mathcal{V}_6^\frac13,
\end{aligned}
\end{equation}
where $\mathcal{V}_6\simeq (-i (T-\bar T))^3$ is the volume of the isotropic six-torus.
Thus, we can write
\begin{equation}
\Lambda_{sp} = \frac{M_P}{\sqrt{\mathcal{S}_{sp}}}  > \frac{M_P}{\mathcal{V}_6^\frac16} =M_s
\end{equation}
where in the last step we identified the string scale $M_s = \mathcal{V}_2^{-\frac12}M_P$ for a compactification on a $k=2$-dimensional manifold with volume $\mathcal{V}_2 = \mathcal{V}_6^\frac13$, down to $d=4$ non-compact dimensions. We see that the species scale is bigger than the string scale. Taylor expanding, we can read off the additive logarithmic correction responsible for this effect
\begin{equation}
\Lambda_{sp} \simeq M_s \left(1+\frac32 \mathcal{V}_6^{-\frac13}\log \mathcal{V}_6^\frac13\right).
\end{equation}

\subsection{String states}

We can follow a similar logic for the dependence of the species scale on the axio-dilaton of the six-dimensional toroidal orbifolds we have been considering.

Towers of string states are governed by the string coupling.
In the asymptotic regions of the moduli space, the string coupling is related to the four-dimensional dilaton $S$ by $-i(S-\bar S) \simeq g_s^{-2}$. Therefore, we have
\begin{equation}
\label{Ngsbndr}
\mathcal{S}_{sp} \simeq g_s^{-2} \simeq (-i(S-\bar S))
\end{equation}
at the boundary of the moduli space. As for the volume modulus, we want to find a duality invariant function with the same asymptotic behaviour for ${\rm Im}\,S\to 0$ and ${\rm Im}\, S \to \infty$. Duality transformations are once more ${\rm PSL}(2,\mathbb{Z})$ transformations and thus $\eta(S)$ is the desired function. We then arrive at 
\begin{equation}
\mathcal{S}_{sp} \simeq- \log |\eta(S)|^{12},
\end{equation}
where the exponent is fixed to match with the asymptotic expression \eqref{Ngsbndr} and we are neglecting factors of $\pi$. The modular invariant completion is then
\begin{equation}
\mathcal{S}_{sp} \simeq -\log \left[(-i(S-\bar S))^3|\eta(S)|^{12}\right].
\end{equation}
For ${\rm Im}S\to \infty$, this behaves as
\begin{equation}
\begin{aligned}
\mathcal{S}_{sp} \simeq g_s^{-2} - 3 \log g_s^{-2},
\end{aligned}
\end{equation}
and thus
\begin{equation}
\Lambda_{sp} = \frac{M_P}{\sqrt{\mathcal{S}_{sp}}} > g_s M_P =M_s,
\end{equation}
where we used the relation $M_s = g_s M_P$. Thus, once more the presence of logarithmic corrections stemming from the Quillen anomaly is responsible for the fact that the species scale is slightly larger than the string scale.
Taylor expanding, we can read off the additive logarithmic correction responsible for this effect
\begin{equation}
\Lambda_{sp} \simeq M_s \left(1+\frac32 g_s^{2}\log g_s^{-2}\right).
\end{equation}
It is not clear at present whether or not the species scale can also receive multiplicative logarithmic corrections associated to string states. Our analysis from the topological free energy and modular invariance suggests that this is not the case, but ultimately the question remains open.

\section{Modular invariant superpotential and swampland conjectures}
\label{sec:potential}

In the previous sections, we gave evidence that the general expression \eqref{F1Torbifolds} captures the moduli dependence of $\mathcal{S}_{sp}$, and in turn of the species scale, in certain toroidal orbifolds preserving four supercharges. In the present section, we connect this function to other quantities of $\mathcal{N}=1$ supergravity, such as the gravitino mass and the scalar potential. Finally, we point out that the anti-de Sitter distance conjecture \cite{Lust:2019zwm} and the gravitino conjecture \cite{Cribiori:2021gbf,Castellano:2021yye} are automatically satisifed in these models, with an exponential decay in terms of the number of species when going towards the boundary of moduli space. Throughout this section, we work in Planck units.

We want to argue that the species scale studied so far is closely related to a well-known quantity in supergravity effective actions, namely the gravitino mass.
The starting point is once more \eqref{F1Torbifolds}, which we rewrite using \eqref{eKoutdet} as
\begin{equation}
\label{eFgravmass}
e^F\simeq e^{-K}\det |W_{\alpha\beta}|^2. 
\end{equation}
One can notice that this expression shares some similarities with that of the gravitino mass in $\mathcal{N}=1$ supergravity, namely
\begin{equation}
m_{3/2}^2 = e^G = e^K W \bar W.
\end{equation}
In fact, the analogy becomes exact if the equation
\begin{equation}
\label{W=1/detW}
W=\left(\det W_{\alpha\beta} \right)^{-1}
\end{equation}
admits non-trivial solutions (we exclude the case $W=0$). We recall that $W_{\alpha\beta}$ is an infinite-dimensional matrix. One can check that the expression
\begin{equation}
\label{Wprodeta}
W(T)=\frac{1}{\prod_{i=1}^3 \eta(T_i)^2} 
\end{equation}
solves \eqref{W=1/detW}. Indeed, since in our models one can compute explicitly that
\begin{equation}
\log \det W_{\alpha\beta}=\sum_{i=1}^3\sum_{{(m_i,n_i)}\neq (0,0)} \log(m_i+n_iT_i),
\end{equation}
the relation \eqref{W=1/detW} becomes
\begin{equation}
 \log W(T_i)= - \sum_{i=1}^3  \sum_{(m_i,n_i)\neq (0,0)}  \log \left(m_i+n_i T_i\right),
\end{equation}
and using the regularization \eqref{sumlogeta} one arrives at \eqref{Wprodeta}.
The intuition behind \eqref{Wprodeta} can be explained as follows. One starts from the holomorphic function $W= const$, which captures the pure field theoretical, but not topological, contribution. To promote it to a function valid over all moduli space, one needs to include the topological contributions, which eventually give a modular invariant expression. A scalar function that has the appropriate moduli dependence, is holomorphic and duality covariant turns out to be $\det W_{\alpha\beta}$. The correct power between $W$ and $\det W_{\alpha\beta}$ can be fixed either by asking that $W$ has modular weight $-3$, or equivalently by demanding that  \eqref{eFgravmass} is the inverse of the gravitino square mass. 

The picture can be generalized further. Given a modular invariant function $H(T)$ and considering the isotropic case for simplicity, $T_i \equiv T$, with similar steps one can see that the equation
\begin{equation}
W = H(\det{W}_{\alpha\beta})^{-1}
\end{equation}
is solved by 
\begin{equation}
W(T) = \frac{H(T)}{\eta^6(T)}.
\end{equation}
This captures the case in which the number of species \eqref{NspT62bu} is increased by
\begin{equation}
\mathcal{S}_{sp} \to \mathcal{S}_{sp} -\log |H|^2,
\end{equation}
namely when there is an additive and modular invariant correction to the topological free energy. Indeed, the topological string free energy is an index-like quantity and, as such, it is defined up to an additive constant, which can here be taken into account by the modular invariant function $H(T)$. This effect is relevant when counting the number of species at the so called desert point of the moduli space \cite{Long:2021jlv}, where the contribution from the topological free energy is minimized. 
As explained in \cite{Cvetic:1991qm} and reference therein, the most general form of $H(T)$ which ensures the absence of singularity in the fundamental domain is
\begin{equation}
H(T) =\left( \frac{G_6(T)}{\eta(T)^{12}}\right)^m\left( \frac{G_4(T)}{\eta(T)^{8}}\right)^n \mathcal{P}(j) = (j-12^3)^\frac{m}{2} j^{\frac n3}\mathcal{P}(j),
\end{equation}
where $m$, $n$ are non-negative integers, $G_4(T)$ and $G_6(T)$ are Eisenstein series defined in \eqref{G2rEisenstein} and $\mathcal{P}(j)$ is any polynomial in the $j$-invariant \eqref{jinv}. Notice that we are neglecting an overall constant when passing from one expression to the other.
Yet a further generalization is to include a dilaton dependence via a function $\Omega(S)$ entering as
\begin{equation}
\label{WST}
W = \frac{\Omega(S) H(T)}{\eta(T)^6}.
\end{equation}
In this more general setup with dilaton $S$ and volume modulus $T$, the K\"ahler potential is given by
\begin{equation}
\label{KST}
K=-\log(-i(S-\bar S)) - 3 \log (-i(T-\bar T)).
\end{equation}

We argued that, in the models we have been considering, the dependence on the moduli $T$ and $S$ of topological free energy and, in turn, of the species scale is in fact captured by the gravitino mass
\begin{equation}
e^F \simeq e^{-G} = \frac{e^{-K}}{W\bar W} \equiv \frac{1}{m_{3/2}^2},
\end{equation}
where $W$ is given by \eqref{WST} while $K$ by \eqref{KST}. Equivalently, we have
\begin{equation}
\label{Nspm32}
\mathcal{S}_{sp} \simeq F\simeq -\log m_{3/2}^2,
\end{equation}
where the minus sign indicates again that we are dealing with the bosonic free energy. Notice that the sign ensures that $S_{sp}\geq 0$, since we work in Planck units.

Having discussed the gravitino mass, the next natural quantity to look at is the supergravity scalar potential
\begin{equation}
V=e^K\left(g^{i\bar\jmath} D_i W\bar D_{\bar\jmath}\bar W- 3 W\bar W\right) = e^G(g^{i\bar\jmath}G_i G_{\bar\jmath} -3).
\end{equation}
Since the K\"ahler metric is positive definite, we have
\begin{equation}
\label{m32>mV}
m_{3/2}^2 \geq -\frac{V}{3},
\end{equation}
which is trivially satisfied for $V\geq 0$ but it becomes more interesting when $V<0$. In this second case, we can recast \eqref{Nspm32} as
\begin{equation}
\label{NFlogV}
\mathcal{S}_{sp}  \leq - \log\left(-\frac V3\right).
\end{equation}
This is an upper bound on the species entropy valid for $V<0$ (trivial for $V=0$)  and following from the supersymmetry algebra.\footnote{A bold statement would be that the bound \eqref{NFlogV} has to be taken seriously for any sign of $V$ and the fact that it leads to an imaginary entropy for de Sitter signals an inconsistency of such vacua.} Indeed, up to numerical coefficients inside the logarithm, a relation such as \eqref{m32>mV} is valid in general in (gauged) supergravity in any number of spacetime dimensions and preserved supercharges. This is because the structure of the supergravity scalar potential is given by the difference between the square of the spin-1/2 supersymmetry variations and that of the spin-3/2 supersymmetry variations, i.e.~schematically $V = (\delta( \text{spin-1/2}))^2-(\delta( \text{spin-3/2}))^2$. Since the last term is the (trace of) the gravitino mass (matrix), one has schematically $m_{3/2}^2 + V = (\delta (\text{spin-1/2}))^2 \geq 0$ in any dimensions and for any preserved supercharges. 

Let us make a couple of comments. First, the bound \eqref{NFlogV} is suggestive of a possible black hole interpretation. Consider a black hole of size $L$ in anti-de Sitter with $L\leq L_{AdS}$, namely fitting within the anti-de Sitter spacetime with radius $L_{AdS}$. By definition, the smallest possible black hole is such that $L^2\simeq \mathcal{S}_p$, in Planck units (and in four dimensions, but the generalization is straightforward). Thus, we get that for such a black hole $\mathcal{S}_{sp} \leq L_{AdS}^2 \simeq (-V/3)^{-1}$. This is close to \eqref{NFlogV} but not quite the same bound. Indeed, the difference between the two is that $(-V/3)^{-1}$ is replaced by the one-loop result $\log (-V/3)^{-1}$. It would be interesting to understand if the discrepancy can be explained in terms of the emergence proposal.  
Second, a bound on $\mathcal{S}_{sp}'$ can be derived by combining \eqref{NFlogV} with \eqref{dLambdaspbound}. Rewriting the latter as
\begin{equation}
\label{Ssp'b1}
(\mathcal{S}_{sp}')^2 \leq (d-2)^2 c_{sp} \mathcal{S}_{sp}^2 ,
\end{equation}
we have then
\begin{equation}
(\mathcal{S}_{sp}')^2 \leq  (d-2)^2 c_{sp}\left(-\log \left(\gamma V\right)\right)^2 ,
\end{equation}
where $\gamma=-\frac13$ in four-dimensional $\mathcal{N}=1$ supergravity but it is different otherwise. 
Thus, in models with a non-trivial, negative scalar potential, the slope of the species entropy is upper bounded by the profile of (the logarithm of) $V$. This bound is valid everywhere on the field space, as long as one uses duality invariant functions of the scalars.

Finally, we can make a direct connection  between the moduli-dependent species scale studied in this note and certain swampland conjectures, namely the anti-de Sitter distance conjecture \cite{Lust:2019zwm} and the gravitino conjecture \cite{Cribiori:2021gbf,Castellano:2021yye}. From \eqref{NFlogV}, we have directly that (recall $\gamma <0$)
\begin{equation}
e^{-\mathcal{S}_{sp}} \geq \gamma V \geq 0. 
\end{equation}
We thus see that the profile of the scalar potential is upper bounded by a decaying exponential. The first inequality is saturated in the supersymmetric case, as it is clear from \eqref{Nspm32}. The above relation is valid over all scalar field space, as long as one uses modular invariant functions. To proceed with a purely bottom-up reasoning, we recall the second law of species thermodynamics proposed in \cite{Cribiori:2023ffn}, which states that when moving towards the boundary of the moduli space, the species entropy cannot decrease. If $\mathcal{S}_{sp}$ remains constant all the way to the boundary, we cannot argue any further. However, we believe this case to be non-generic from a swampland perspective, since the number of species is expected to increase when approaching asymptotic regions, as reviewed in section \ref{sec:dc}. On the other hand, if $\mathcal{S}_{sp}$ increases, we see that the cosmological constant has to decrease at least exponentially with the species entropy. In the supersymmetry case, the exponential decay is exact, while in the non-supersymmetric case our discussion does not exclude a faster decay. Let us stress that this exponential behaviour is not an assumption: it follows from the very definition of free energy as the logarithm of the partition function. Thus, we understand that in these models the gravitino conjecture and the anti-de Sitter distance conjecture on supersymmetric vacua are always satisfied, since the limit $m_{3/2}\to 0$ leads to $\mathcal{S}_{sp}\to \infty$ and thus to $\Lambda_{sp} \to 0$.\footnote{A similar statement for the one modulus model with the simple superpotential $W=1/\eta(T)^6$ appeared already in \cite{Cribiori:2022sxf}.} The states predicted by these conjectures are precisely the species states becoming light.

\section{Discussion}
\label{sec:discussion}

In this note, we pointed out that modular invariant superpotentials and scalar potentials of $\mathcal{N}=1$ supergravity provide a moduli dependent expression of the species scale. Building on the work \cite{Ferrara:1991uz},  we explained how this expression can be derived from the topological free energy of the $E_8\times E_8$ heterotic string compactified on toroidal orbifolds. Then, we showed how the moduli dependence can be generalized by introducing further modular invariant functions of the scalar fields. We observed that in asymptotic regions of the moduli space, especially at large volume and weak string coupling, the species scale receives additive logarithmic corrections in such a way that the resulting value is slightly bigger than the string scale. On the other hand, we could not find any multiplicative logarithmic correction, whose existence remains thus an open problem. Finally, we pointed out that the anti-de Sitter distance conjecture and the gravitino conjecture of the swampland program are automatically satisfied in the setups here investigated. 

The present work can be extended along several directions. For example, it would be interesting to consider the moduli dependence of more complicated Calabi-Yau manifolds and repeat our discussion by replacing modular invariance with mirror symmetry.  
In general, it would be important to understand better the relation between the topological string free energy and the prepotential of the supergravity effective action. While there is large amount of evidence that a qualitative relation exists, which is a necessary condition for the work here presented to be well-motivated, the precise details of this relation are still elusive \cite{Cardoso:2008fr, Cardoso:2010gc, Cardoso:2014kwa}. As for the models here investigated, the results of \cite{Cardoso:2008fr} suggest that at one loop this issue does not arise, since the genus-one topological string free energy and the first higher derivative correction to the supergravity effective action do match. However, an explicitly discrepancy is found at two loops in certain classes of models. Thus, at least at a conceptual level, this raises the question on what one should look at to define the species scale, namely if the topological string free energy or the supergravity effective action. We hope to come back to this problem in the future.

\paragraph{Acknowledgments.}
We thank R.~Blumenhagen, A.~Gligovic, A.~Paraske\-vo\-poulou, N.~Righi and M.~Scalisi for discussions. 
N.C.~thanks the group at DESY for hospitality while this work was under completion. The work of N.C.~is supported by the Alexander-von-Humboldt foundation.
The work of D.L.~is supported by the Origins Excellence Cluster and by the German-Israel-Project (DIP) on Holography and the Swampland.

\appendix

\section{Conventions on modular functions}
\label{app:conv}

Here, we collect some useful formulas and conventions concerning modular functions. A recent and comprehensive review on the topic with applications to string theory can be found in \cite{DHoker:2022dxx}.

In our conventions, the Dedekind eta-function is defined as
\begin{equation}
\eta(T) = e^{\frac{i\pi }{12}T} \prod_{n=1}^\infty \left(1-e^{2\pi i n T}\right)
\end{equation}
and under ${\rm SL}(2,\mathbb{Z})$ generators it transforms as
\begin{equation}
\eta(T+1) = e^{\frac{i\pi }{12}T} \eta(T),\qquad \eta\left(-\frac{1}{T}\right) = \sqrt{-i T}\eta(T).
\end{equation}
Furthermore,  one has
\begin{equation}
\log\eta(T) =-\frac{\pi}{12}  {\rm Im}T+ \mathcal{O}(e^{-2\pi {\rm Im}T}),
\end{equation}
implying 
\begin{align}
    \eta(T) &= e^{-\frac{\pi}{12} {\rm Im}T} +\dots, \qquad\qquad\qquad\qquad \,\text{for}\quad {\rm Im}T\to \infty,\\
    \eta(T) &= \sqrt{{\rm Im}T}e^{-\frac{\pi}{12} {\rm Im}T}+\dots, \,\,\quad\qquad\qquad \,\text{for}\quad {\rm Im}T\to 0,
\end{align}
where dots stand for terms suppressed in the limit.

We define the Eisenstein series
\begin{equation}
\label{G2rEisenstein}
G_{2r}(T) = \sum_{(m,n)\neq(0,0)} \frac{1}{(m+nT)^r},
\end{equation}
where the sum is over all values except $m=n=0$. This is absolutely convergent for $r>1$, but only conditionally convergent for $r=1$. As a consequence, depending on which regularization one chooses one can preserve either holomorphy or modularity. The regularized holomorphic series is 
\begin{equation}
G_2(T) = -4\pi i \frac{\partial \log\eta(T)}{\partial T},
\end{equation}
such that
\begin{equation}
G_2\left(\frac{aT+b}{cT+d}\right) = (cT +d)^2 G_(T) -2\pi i c (cT +d).
\end{equation}
We see that there is an additional piece which spoils modular transformations. The regularized series transforming correctly under modular transformations is non-holomorphic and defined as
\begin{equation}
\hat G_2 (T,\bar T) = G_2(T) - \frac{\pi}{{\rm Im}T}.
\end{equation}
Another useful function is the $j$-invariant, which can be given in terms of the Eisenstein series as 
\begin{equation}
\label{jinv}
j(T) = \frac{(12 g_2)^3}{\Delta},
\end{equation}
where $g_2(T)=60 G_4(T)$, $g_3(T)=140 G_6(T)$ and $\Delta = g_2^3-27 g_3^2$.

Next, we recall two results from Riemann zeta function regularization following the appendix A of \cite{Ferrara:1991uz}. The Riemann zeta function is defined as
\begin{equation}
\zeta(s) = \sum_{n=1}^\infty n^{-s}
\end{equation}
and we will need the special values $\zeta(0)=-\frac12$, $\zeta(-1)=-\frac{1}{12}$. Then, one can prove
\begin{align}
\label{2xsum1}
\sum_{(m,n)\neq (0,0)} A&= -A,\\
\label{sumlogeta}
    \sum_{(m,n)\neq (0,0)} \log(m+nT) &= 2 \log \eta(T) + \dots,
\end{align}
where dots stand for terms not depending on $T$, while $A$ is any object not depending on $m,n$.
Notice that from \eqref{2xsum1} one can show that
\begin{equation}
\label{eKoutdet}
F = \log \det \left(e^K |W_{\alpha\beta}|^2\right) = \log \left(e^{-K}\det |W_{\alpha\beta}|^2 \right).
\end{equation}

\bibliography{references}  
\bibliographystyle{utphys}

\end{document}